# From Measuring Land Use Mix to Measuring Land Use Pattern - New Methods for Assessing Land Use

Yan Chen, University of North Carolina at Chapel Hill

Yan Song, University of North Carolina at Chapel Hill

Abstract

Land use mix is one of the central concepts in the urban planning field, though its measure has been found to have many fallacies. In this study, we propose multiple alternative methods to the Conventional Shannon Entropy land use mix index, which is generally employed to measure land use diversity. The study attempted to measure land use mix from two new perspectives and test the performance of the new measures with economic, transportation, and social outcomes. We first offered an improved single measure, entropy-based land use mix index weighted by adding surrounding land use attributes and the regional land use context to mitigate the modifiable area unit problem (MAUP) and equal composition issue. Second, we designed a multi-measure method to describe land use patterns via clustering. We found that the new single measure method is more effective than the existing Conventional Shannon Entropy index in accurately delivering the vision of land use mix that Jane Jacobs originally proposed. We also found the multi-measure clustering method performed much better in the regression model on various effect variables, compared with both conventional and other new index. Upon further review, we found that land use pattern complexity's relationship with various outcomes makes a single definition measure very difficult to represent true land use fully. Therefore, we recommend that future research measure whether or not land use is mixed to then assess land use pattern.

Keywords: Land use mix, land use mix measure, built environment, spatial analysis

# 1. Introduction

In 1961, Jane Jacobs said, "Intricate mingling of different uses in cities are not a form of chaos. On the contrary, they represent a complex and highly developed form of order." Since Jacobs (1961) first raised the importance of land use mix in urban planning, the diversity in land use has become a central target of research and practice on built environments regarding urban planning, transportation planning, health and preventive medicine, and property valuation (Ewing & Cervero, 2010; Frank et al., 2005; Hess et al., 2001; Song & Knaap, 2004). The benefits of land use mix have been proved in several fields, most notably in place-making, transportation, public health, and urban economics. From the place-making point of view, mixed uses would increase the vibrancy of public places such as streets, plazas, and parks by offering more opportunities for different activities. From a transportation perspective, the advantages of mixed uses would bring various places of origins and destinations closer to each other; therefore, encourage non-motorized modes and shorter travel distances. From a public health perspective, bringing a variety of desirable destinations closer to residential areas will promote walking travel modes and physical activities, thus improve human physiological conditions. Moreover, from an economic point of view, the proper land use mix is expected to increase land value and boost higher density development through the support of urban amenities (Song & Knaap, 2004). Therefore, mixed land use has become one of the central concepts in the urban planning field today. For example, Calthorpe (1993) emphasized that "all TODs (Transit-Oriented Developments) must be mixed use and contain a minimum amount of public, core, commercial and residential use. Vertical mixed-use buildings are encouraged, but considered a bonus to the basic horizontal mix use requirement." Aside from TODs, mixed land use has been mentioned in possibly every planning idea in recent years, such as "smart growth" (Downs, 2005), "new urbanism" (Congress for the New Urbanism, 2000), and "complete neighborhood" (Pivo, 2005).

Therefore, it is essential to develop a widely accepted and valid quantitative measure for the land use mix concept to empirically test its causes and effects with various urban environment and life factors. Through a simulation and comparison of the characteristics of various land use diversity indexes, Song et al. (2013) found that Shannon entropy is the most efficient way of representing land with three or more uses. Therefore, the Shannon entropy–based index has become the most influential and commonly accepted method for measuring land use mix in quantitative urban studies. For instance, in the transportation field, Shannon entropy is adopted in the "6Ds" research framework within the "diversity" concept measure, which has been applied to dozens of studies (Ewing & Cevero, 2010).

Given the importance and popularity of the Shannon entropy measure in urban studies, the validity of this measure should be critical for all empirical studies that apply it to examine the effect of built environments. If researchers or planners cannot estimate the effects of land use diversity in their analysis and planning, all arguments supporting mixed land use may be vulnerable to criticism. However, in recent years, more and more researchers have found that the Shannon entropy–based land use mix index has many deeply rooted issues that heavily jeopardize its construct validity (Im & Choi, 2018; Hess 2001; Manaugh & Kreider, 2013), which means the studies utilizing it may not accurately measure the vision of what the land use mix concept, proposed by Jacobs (1961), was originally trying to deliver.

Therefore, in this study, we review and summarize the current criticisms and limitations of the conventional Shannon entropy (CSE) index from previous research. Then based on this reflection, we propose three improved alternatives to the index by (1) adding geographical weight (GW) to the entropy equation, (2) adding both GW and regional structural weight (RSW) to the entropy equation, and (3) using cluster methods to summarize land use mix pattern based on GW modified land use data from alternative (1).

Are the new measurements able to fix CSE's validity issue in measuring land use mix?

Does the single measure (GW/RSW) or the multi-measure (clustering) method better improve validity regarding different research subjects?

The remainder of this paper is organized as follows: Section 2 reviews the development of the CSE index and summarizes its limitations. Section 3 explains the concepts and generation of GW and RSW and how they were added to CSE to create new indexes. This section also clarifies the clustering method we used to identify U.S. land use mix patterns. Section 4 identifies the performance of CSE, GW, GW & RSW, and GW & clustering in both qualitative and quantitative methods to demonstrate the improved validity of the new indexes. In the final section, we discuss our findings and the contributions of the new land use mix measures.

# 2. Literature Review

## 2.1 Land Use Mix Measure

CSE was originally applied in 1949 to measure the diversity of animal species in environment systems (Shannon & Weaver, 1949). The term *entropy* references an analogy to statistical mechanics borrowed from physics that measures the degree to which different functions occur equally within a given range. After being used in the environmental field for so long, CSE was first introduced in urban studies by Cervero (1989) as a quantitative urban form measure using geographical information systems (GIS). It was later formalized by Cervero and Kockelman (1997) in the "3D" model to become a standard for describing the status of land use mix in transportation studies. After that, CSE's uses extended from transportation-related research to most built environment–related topics in urban studies.

However, the logical principles for using CSE have not yet been specifically set forth.

Cervero (1989), who initially proposed the Shannon entropy index to measure land use diversity, did not sufficiently explain the relationship between the land use diversity index and its impact.

The entropy index varies from 0 to 1 and can be expressed as the equation below:

$$CSE = \frac{-\sum A_{ij} ln A_{ij}}{ln N_j}$$

$A_{ij}$ means the percent of land use i in land parcel j
$N_j$ means the total number of represented land use types

For the interpretation, 1 is the maximum entropy index value that can only be achieved by a perfectly equal land use balance, such as 33%/33%/33%. 0 is the minimum value which means there is only one land use type in the parcel.

## 2.2 Limitations and Criticisms of the CSE Index

## 2.2.1 The modifiable areal unit problem and boundary effect

Both the modifiable areal unit problem (MAUP) and boundary effect have long existed in spatial analysis. Land use mix, as a typical geographical concept, also suffers from these drawbacks. The MAUP refers to study results being influenced by both the shape and scale of the measure aggregation unit, which means that the CSE index calculated at the census block level would be highly different from that calculated at the neighborhood or county level. The boundary effect means that interactions or interdependencies across the borders of the bounded unit are ignored or distorted. In essence, both the MAUP and boundary effect issue originate from one characteristic of land use mix: land use mix is a spatially dependent concept whose performance both the land use attributes within the site unit and its neighbors will influence.

As illustrated in Figure 1, suppose there are four 500-foot street blocks with different land uses for each block. When considered as an aggregated region, this would be highly mixed use; however, when considered individually, each block would be typical single use.

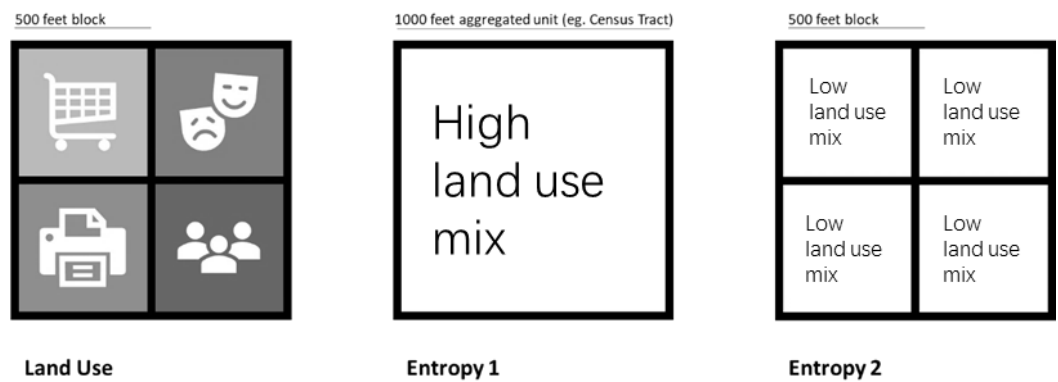

Figure 1. The MAUP issue of land use mix measure

Therefore, for a valid measure of the land use mix index, not only the land use within each block but also the land use attributes surrounding the site must be considered to mitigate the MAUP and boundary effect issues. To address this issue, Hess et.al (2001) developed a flexible boundary method that would group all nearby parcels with functionally complementary land uses. Manaugh and Kreider (2013) developed another interaction method for measuring land use mix that accounts for the extent to which complementary land uses adjoin one another. After testing the new measure's relationships with travel behavior, the results show the usefulness of the new approach in that it significantly improves model fit, compared to the CSE index. Both two papers initiated the improvement of the CSE index; however, the land uses in these studies are oversimplified to three or four categories, which further limits the result's external validity. Hess's study also drops large amount of areas from analysis due to the "non-complementary" land use definition. If research on built environments aims to derive repetitive values to accurately reflect various land use mixes, the index must address sufficient comprehensibility. This would also limit the externality of the measure and the application of that method. In addition, these researchers failed to address other issues of CSE, which are clarified below.

## 2.2.2 Composition and regional context

Another strong criticism of CSE is its definition of best composition. The most proper and feasible mixed use share of a site should be comparable to the share of the regional context instead of an arbitrarily defined equipartition. However, for CSE, the highest index value indicates that every component within it shares an equal proportion. Previous work has referred to this even land use mix as "perfect" (e.g., Rajamani, Bhat et al., 2003). In contrast, other scholars have argued that these studies seem to lack a compelling theoretical underpinning for this key point (Manaugh & Kreider, 2013; Moudon et al., 2001). Land use types are not equally proportional by nature. Usually, residential uses take up the vast majority of lands, and such uses can be permitted for up to 50% of the total area, while other types, such as retail, take up less than 5% of lands in a city. Therefore, forcefully promoting an equal proportion of land use mix within certain sites will inevitably lead to a lack of land use mix in other areas. Using empirical statistical methods, Im and Choi (2018) found that CSE had an inconsistent quadratic relationship with pedestrian volume; they further showed that a less equally distributed land use mix performed better in an area than the perfectly equally distributed land use mix.

Past research has addressed some assumptions made in using the indexes outlined above. For example, Kockelman (1996) was the first to propose standardizing CSE and briefly described its limitations in a note, stating, "The index of land use balance, entropy, remains constant when distinct land use types remain in constant relative proportions; yet mixing or integration of land uses can change dramatically" (p. 49). Hess et,al (2001) identified three main issues with any mixed use measure based solely on the proportions of various uses, which are as follows: 1) land use types are not differentiated—a perfect mix of industrial and park uses may score identically to the same proportions of residential and commercial uses, for instance; 2) a measure of proportion misses any sense of interaction (whether the land use is adjoining or separated by barriers); and 3) CSE is symmetric in terms of land use types. That is, suppose the distribution of three land use types is 60%/25%/15%; this scenario will produce the same entropy index as a distribution of 15%/60%/25%. The change of two scenarios have no effect on CSE score, although the implications of the two patterns could undoubtedly be quite different.

Given the considerations outlined above, it is highly important to define the proper "best" land use mix composition instead of the default "equipartition" based on the regional land use context. Although CSE is trapped by this issue, in the economics field, there is another index specifically designed for such a problem. The location quotient (LQ) is a method used to quantify a site's industrial specialization relative to a larger regional geographic unit (Isserman, 1977). It can reveal what makes a site "unique" compared with the regional average. For example, an LQ of 1.0 in mining would mean that the site and region are equally specialized in mining.

Im and Choi (2019) integrated the LQ into CSE to create a new land use index and demonstrated that this new index can improve the explanatory power of the estimation model for the relationship between pedestrian volume and built environment. However, their method still has two limitations. First, it does not use a fixed definition of region, so it cannot be employed for big datasets with large numbers of observations, such as the data we include in this study. This is because the land use context of a *region* varies in different places; for instance, the land use in an industrial city like Detroit is extremely different from that in a recreational city like Miami. Therefore, a dynamic region definition is necessary for such a method. The second limitation is that Im and Choi's (2018) method does not consider the MAUP and boundary effect described above.

To summarize, although CSE has been used in the urban planning field for more than 30 years, it still has many issues that are deeply rooted in its original equation. Without properly handling them, a huge construct validity issue emerges for all the studies using it. There have been some early attempts to improve the CSE index, but these are not comprehensive enough to cover all its issues. Therefore, in this study, we propose a more complex method to generate a new index that addresses all of CSE's issues.

# 3. Data and Method

## 3.1 Data Source

Due to the lack of consistently formatted, high-resolution land use data within the U.S., in this study, we chose to use employment data from different NAICS (North American Industry Classification System) sectors to gather land use information. The definitoin of classification is shown in Table 1 below. This employment data has been used to indicate land use in urban studies before (Ramsey & Bell, 2014). We believe this indicator data also reflects land use information more accurately compared with parcel data by showing not only the land use area, but also the intensity of each type of use.

Table 1. Land use type definition based on NAICS employment data

| Land use type | NAICS employment sector ID |
| --- | --- |
| **Office** | Information; Finance and Insurance; Real Estate, Rental, and Leasing; Management of Companies and Enterprises; Public Administration; Management of Companies and Enterprises |
| **Retail** | Retail Trade |
| **Industrial** | Agriculture; Forestry; Fishing and Hunting; Mining, Quarrying, and Oil and Gas Extraction; Utilities; Construction; Manufacturing; Wholesale Trade; Transportation and Warehousing |
| **Service** | Professional, Scientific, and Technical Services; Administrative and Support; Waste Management and Remediation Services; Health Care and Social Assistance; Educational Services |
| **Entertainment** | Arts, Entertainment, and Recreation; Accommodation and Food Services |

The employment data were collected from the U.S. Census 2015 LEHD Origin-Destination Employment Statistics dataset. It summarizes U.S. employment at the census block level for all states except Wyoming and Massachusetts. The new entropy measure utilized the LODES Work Area Characteristics tables for employment tabulations. We also used the 5-year American Community Survey (ACS) 2013–2018 data for population information to indicate residential land use. We then summarized six types of land use, which are Office, Retail, Industry, Service, Entertainment, and Residential. All land use measures were aggregated to the census block group level, then transferred into the ratio of total volume for further analysis.

For the dependent variables, housing price and the percentage of people who drive alone to commute to work were also extracted from the ACS dataset. Tweet density was compile from tweet data collected through the Twitter stream API within a whole year of 2017. This allowed us to gain a 1% random sample of all geotagged tweets within the contiguous U.S. Then, tweet density was spatially joined with the census block group to summarize its count. Finally, the tweet density count was normalized by the land area of each census block group to compute density. The summary statistics are shown in Table 2.

**Table 2. Summary statistics**

|  | Statistic | N | Mean | St. Dev. | Min | Max | Source |
|---|---|---|---|---|---|---|---|
| **Land use (represented by employment)** | Retail | 213,438 | 71.5 | 222.8 | 0.0 | 17,596.0 | LEHD |
|  | Office | 213,438 | 91.7 | 703.8 | 0.0 | 111,125.0 | LEHD |
|  | Industry | 213,438 | 151.1 | 619.8 | 0.0 | 47,645.0 | LEHD |
|  | Service | 213,438 | 255.8 | 978.5 | 0.0 | 136,086.0 | LEHD |
|  | Entertainment | 213,438 | 70.2 | 333.5 | 0.0 | 97,084.0 | LEHD |
|  | Residential | 213,438 | 638.3 | 393.3 | 1.0 | 18,594.0 | LEHD |
| **Dependent variables** | Housing price | 174,192 | 1,097.5 | 504.5 | 99.0 | 3,501.0 | ACS |
| | Commuting – Drive alone | 212,905 | 0.8 | 0.2 | 0.0 | 1.0 | ACS |
| | Tweet density | 213,438 | 3.7 | 51.5 | 0.0 | 14,623.9 | A* |

* A indicates that the data was collected by the author

## 3.2 New Entropy Equation

To address the issue of traditional CSE measure that we discussed before, we proposed multiple methods to mitigate the MAUP and better reflect the regional context of land use. As shown in Figure 2, the most important components of the new measure are the two spatial weights added to the equation: GW and RSW, which we now explain further.

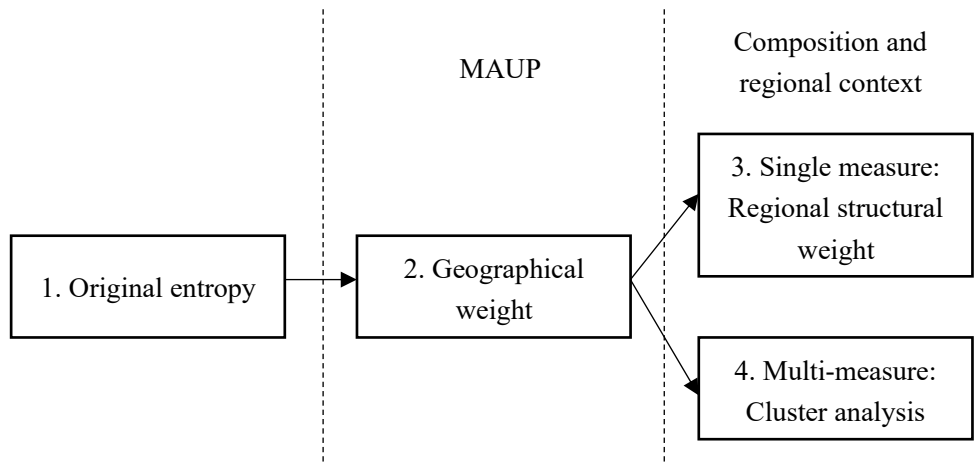

Figure 2. The structure of new land use mix measures

## 3.3 Geographical Weight

GW is designed to address the MAUP. The GW equation can be written as:

$$WL_{mi} = \sum_{j \sim i} LUV_j \times WG_{ij}$$

$$WG_{ij} = 1 - \frac{D_{ij}^2}{D_{max}^2}$$

- $WL_{mi}$ = Sum of locally weighted land use volume of type m
- j~i = All parcels within the local buffer range from the site
- $LUV_j$ = Land use volume for type m within site j
- $WG_{ij}$ = Local weight for site j based on its distance from site i; 0 indicates the edge of the range, while 1 is the center of the range
- $D_{ij}$ = Distance between parcel i and parcel j
- $D_{max}$ = Distance range set for the local buffer

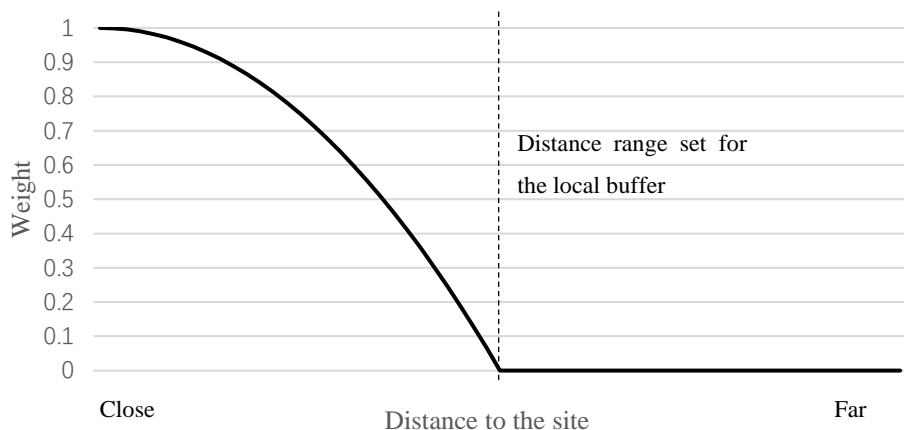

Figure 2. The change of geographical weight by distance to site

The GW allows the measure not only to consider the land use volume within a site, but also to the land use of the surrounding area within a proper distance. However, simply adding all the employments of the site's surrounding area ignores another essential fact: the land use of the nearby regions clearly will have a more substantial influence on the site compared with far away regions. Therefore, we added a local GW for each part of the surrounding area in the equation and then performed a spatially weighted sum for the local area's employment. As shown in the figure 3, the weight will drop by distance to the site from 1 to 0 at the range we set for the max search buffer range. For this weight, 0 indicates the parcel located at the edge of the range, while 1 refers to the parcel located at the center of the range (the site itself). The equation of the weight is derived from the geographically weighted regression that focuses on a similar issue in spatial analysis (Fotheringham et.al, 2003).

The procedure of creating this GW was first to set a buffer range to define the "local" range for all the land parcels. In this study, we set 0.5 miles as the range, which included the surrounding area proximal by 10–15 minutes walking. We used walking in our criteria, as we believe this follows Jacobs's (1961) initial intention of creating a "land use mix" concept.

After setting the buffer for each parcel, we selected the parcels, of which the centroids are within the buffer range, and calculated the geographical spatial weight based on the distance between the surrounding parcel and central parcel. This helped minimize the influence of the edge parcel land use volume on the weighted sum of total land use volume. Figure 4 illustrates an example of local weight generations: assuming the original land use value is 3, with three blocks of 1, 1, and 2 surrounding it, and based on the GW, the final land use value is actually 3.93.

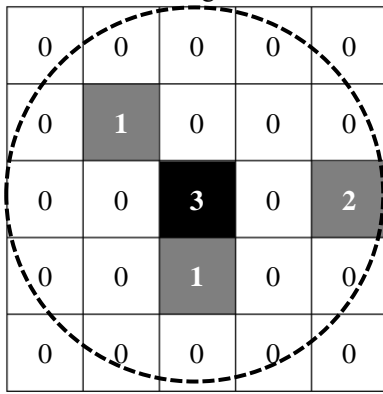
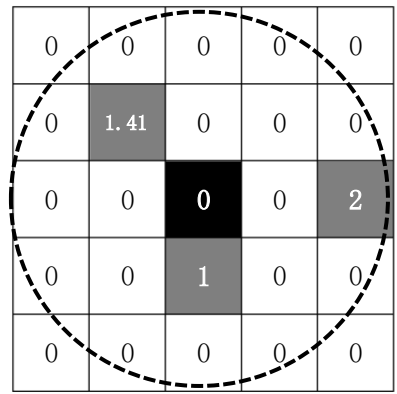
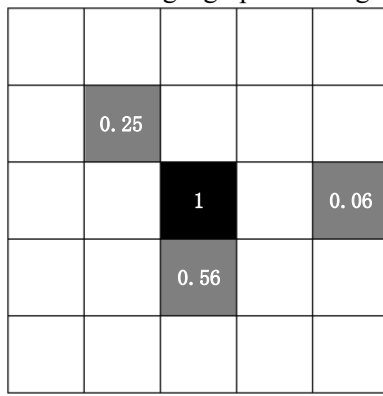

Figure 4. The calculation of geographical weight

## 3.4 Regional Structural Weight

For land use components and regional context issues, we created an RSW to adjust entropy by its regional land use context. To fit the theory paradigm of complete community, the new regional weighted entropy reflects how land use in a site is similar to regional land use composition, instead of showing if all land use occurs in equal percentages. For instance, the figure 5 below presents two land use parcels with different patterns inside. The blue parcel has an equal portion for all land use types, while the orange parcel is associated with the regional context, so residential use is much higher than the remaining uses. CSE would think the blue parcel has a stronger land use mix, but after adjusting with RSW, the result instead validates the orange parcel as having a better land use mix.

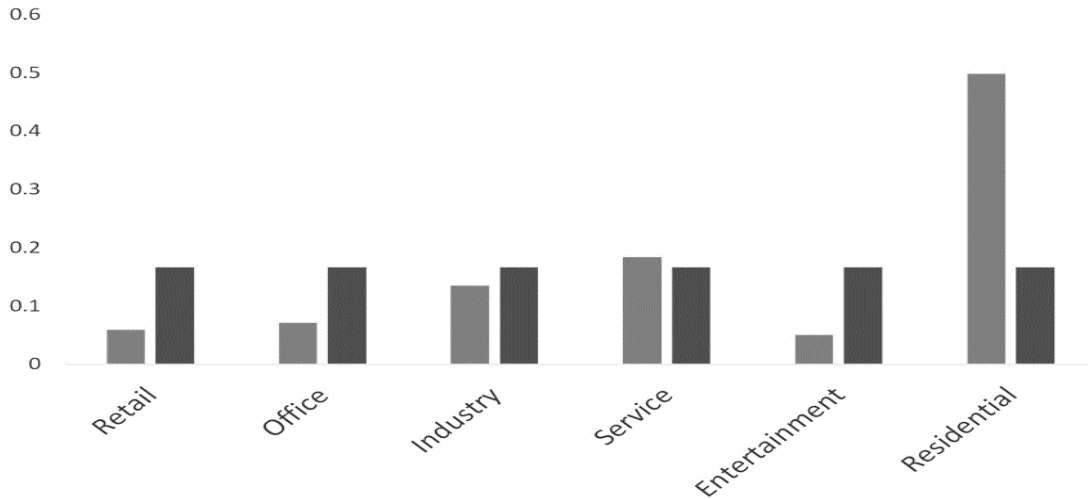

Figure 5. The example of two types of land use pattern

To derive the weights, we first created a regional buffer and identified all the land parcels within it. The regional buffer's range was set to 100 miles to contain most areas in the region. Then, we added up the total land use volume in the region by each type of land use. Finally, we divided the total land use volume to get the regional weight for each type of land use. This technique was also applied to the LQ equation, which similarly addressed this regional context issue, but has been rarely used in land use analysis before:

$$WR_{mi} = \sum_{k \sim i} \frac{Tot_{mik}}{LUV_{mik}}$$

- m = Type of land use
- i = Site of calculated land use mix entropy
- k~i = All parcels within the regional buffer range from the site
- $Tot_{mik}$ = Total land use volume for all types of employment

After combining RSW with GW in the entropy equation, the final entropy index equation can be written as:

$$E = \frac{\sum_{m=1}^{n} \frac{W_{mi}}{W_{toti}} \times \ln(\frac{W_{mi}}{W_{toti}})}{-\ln(n)}$$

$$W_{mi} = WR_{mi} \times WL_{mi}$$

$$WL_{mi} = \sum_{j \sim i} LUV_j \times WG_{ij}$$

$$WR_{mi} = \sum_{k \sim i} \frac{Tot_{mik}}{LUV_{mik}}$$

$$WG_{ij} = 1 - \frac{D_{ij}^2}{D_{max}^2}$$

$$W_{toti} = \sum_{m=1}^{n} Wmi$$

- $E$ = New entropy created based on regional and local weight
- i = Site of calculated land use mix entropy
- m = Type of land use
- n = Total number of land use types
- $W_{mi}$ = Weighted land use volume for type m
- $W_{toti}$ = Sum of total weighted land use volume of all types
- $WL_{mi}$ = Sum of locally weighted land use volume of type m
- j~i = All parcels within the local buffer range from the site
- $LUV_j$ = Land use volume for type m within site j
- $WG_{ij}$ = Local weight for site j based on its distance from site i; 0 indicates the edge of the range, while 1 refers to the center of the range
- $D_{ij}$ = Distance between parcel i and parcel j
- $D_{max}$ = Distance range set for the local buffer
- $WR_{mi}$ = Sum of total land use volume for type m within the regional buffer's range
- k~i = All parcels within the regional buffer range from the site
- $Tot_{mik}$ = Total land use volume for all types of employment

## 3.5 Clustering Methods

Cluster analysis is a technique that combines observations into groups based on their similarity within a set of predetermined characteristics. In this study, it is performed to identify a built environment pattern. In specific, the "K-means" cluster analysis was selected used to classify all census block groups into pattern groups based on similar and dissimilar land use features derived from the previous step. Each land use group is expected to be internally as similar as possible but externally dissimilar to other land use groups.

The K-means clustering began by grouping observations into a predefined number of clusters. It evaluated each observation and moved it to its nearest cluster. The nearest cluster was the one with the smallest Euclidean distance between the cluster's observation and centroid. When a cluster changed after losing or gaining an observation, the cluster centroid was recalculated. In the end, all observations ended up in their nearest cluster. The number of expected clusters must be preset for K-means clustering and was decided via the Elbow method. Elbow method is to locate a balanced point between number of cluster and total within variance by finding the turning "elbow" point. As shown in the figure 6, we decided to cluster the land use composition into 10 groups at using Elbow method.

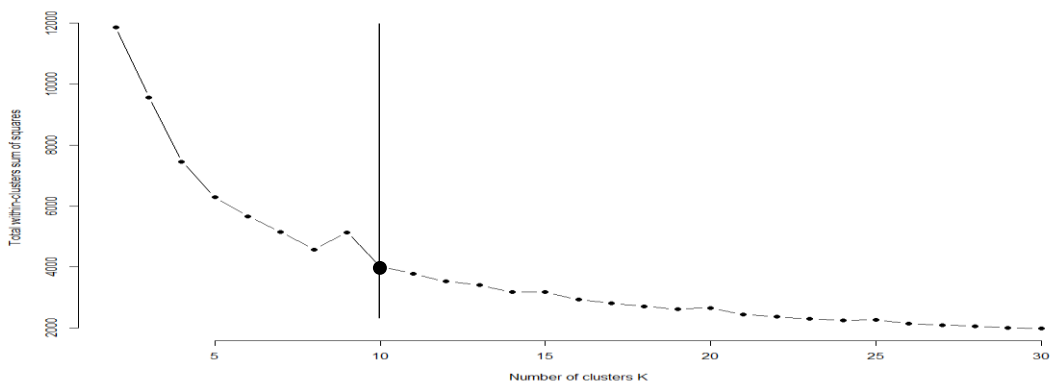

Figure 6. The Elbow method for cluster analysis

## 3.6 Comparison Methods

In this paper, we intended to demonstrate the advantages of the new measurement methods. Therefore, both qualitative case study and quantitative regression methods were used to illustrate the advantages of the new methods. For the qualitative method, we selected two sites in the U.S. and compared their CSE, GW, and GW & RSW measures. Then, with the help of Google Maps, we explored the ground face of the sites' land use mix situation by looking at the buildings and facilities both inside and around the sites to see which index better reflected real land-use mix.

For the quantitative methods part, it's essential to maintain the clarity and interpretability of the result for the complex land use mix effects inside the city. Therefore, we decided to take the parsimonious approach to test the measure's performance in several most represented topics of land use mix rather than an exhaustive list of the variables. Three primary outcomes are selected in this study:

(1) housing price, as a representation of the economic influence

(2) percentage of people who drive alone to commute, as a representation of transportation influence

(3) tweet density, as a representation of social vibrancy influence.

Also, we chose linear regression, one of the most commonly employed empirical study methods in urban planning research, to examine the associations between land use mix and its effects. To compare the indexes' performance, three critical indicators from the regression results are extracted:

(1) coefficient, to see if the direction meets the previous theories

(2) t-value, to compare the level of significance for the results

(3) R-square, to compare the indexes' explanatory power

# 4. Results

## 4.1 Case Study of Index Changes

There are two cases we followed. The first, from New York, illustrates the importance of GW, and the second case, from Chicago, shows the importance of using both GW and RSW. For the convenience of interpretation, each land use type was transferred to its portion of the total volume in the text and table below.

The New York case is located between 29$^{th}$ and 30$^{th}$ Street and 8$^{th}$ and 9$^{th}$ Avenue. Its geocode is 36061009700. The table 3 shows that this census block group itself is mainly residential area, which occupies 86.15% of its land use; the original CSE is only 0.34. However, the surrounding areas of this block group include many other different land uses, such as a New York state office in the north, Penn Station in the west, and many restaurants and bars along 8$^{th}$ Avenue. Therefore, after taking these into account using GW, the entropy rises to 0.90, which undoubtedly better reflects the real situation of land use mix on-site.

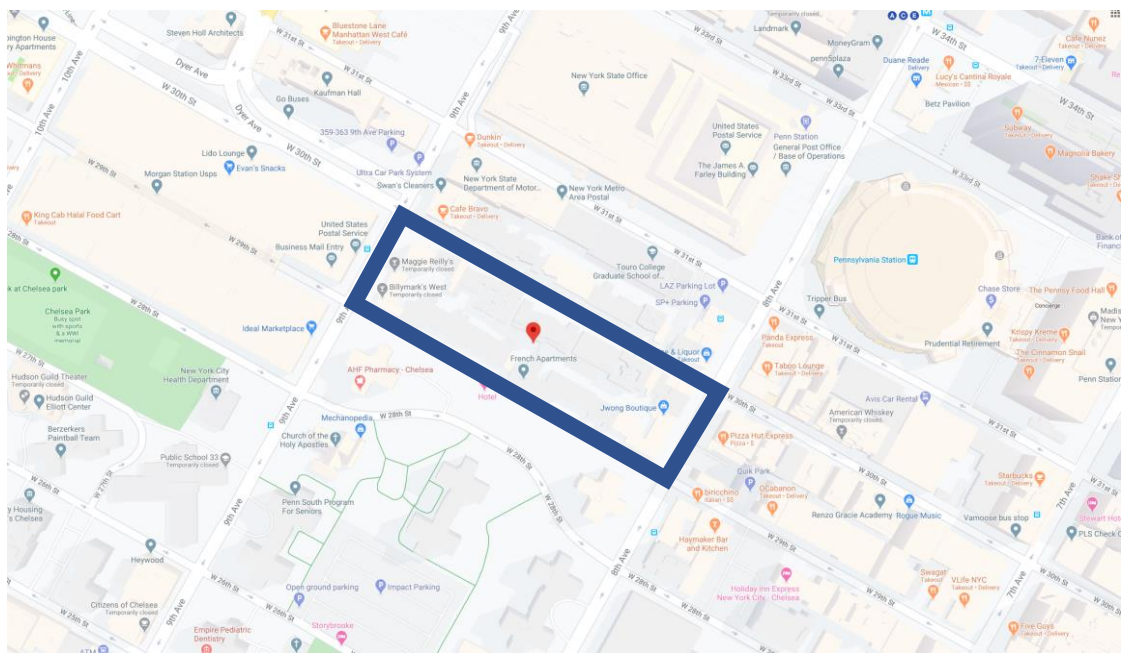

**Figure 7 The map of New York case**

**Table 3. The entropy of CSE and GW modified index for the New York case**

|  | Retail | Office | Industry | Service | Entertainment | Residential | Entropy |
|---|---|---|---|---|---|---|---|
| **Original CSE** | 1.11% | 2.39% | 2.23% | 4.14% | 3.98% | 86.15% | **0.34** |
| **+ GW** | 5.93% | 17.44% | 10.02% | 23.28% | 7.83% | 35.51% | **0.90** |

The Chicago case is located in the southern part of the city near West 31st Street. The Geocode is 170316009003. Similarly, this census block group by itself has very concentrated residential use at 93.5% of total volume. The original CSE is only 0.16. However, the map shows that there are two local main streets—South Halsted Street and West 35th Street—located at only 200 meters east and south of the site that are full of restaurants and bars. There is also a church right outside the block's eastern border. So, using GW, we can see the proportion of retail, office, industry, and service use highly increases after considering the surrounding areas. The entropy increases as well to 0.53, which indicates a median land use mix. The reason why this case's entropy is not higher is that residential use is still at 76.2% of total land use volume. However, as we mentioned before, residential use is supposed to take a more substantial proportion of land use compared with other land use types, not an equal proportion, as the CSE measure designed. Therefore, we used a buffer to extract the land use structure within a 100-mile radius around the site, which helped define the "ideal" land use mix proportion for the site. As expected, the issue of high residential use is mitigated after using RSW, and the entropy further increases to 0.90, which indicates a good land use mix.

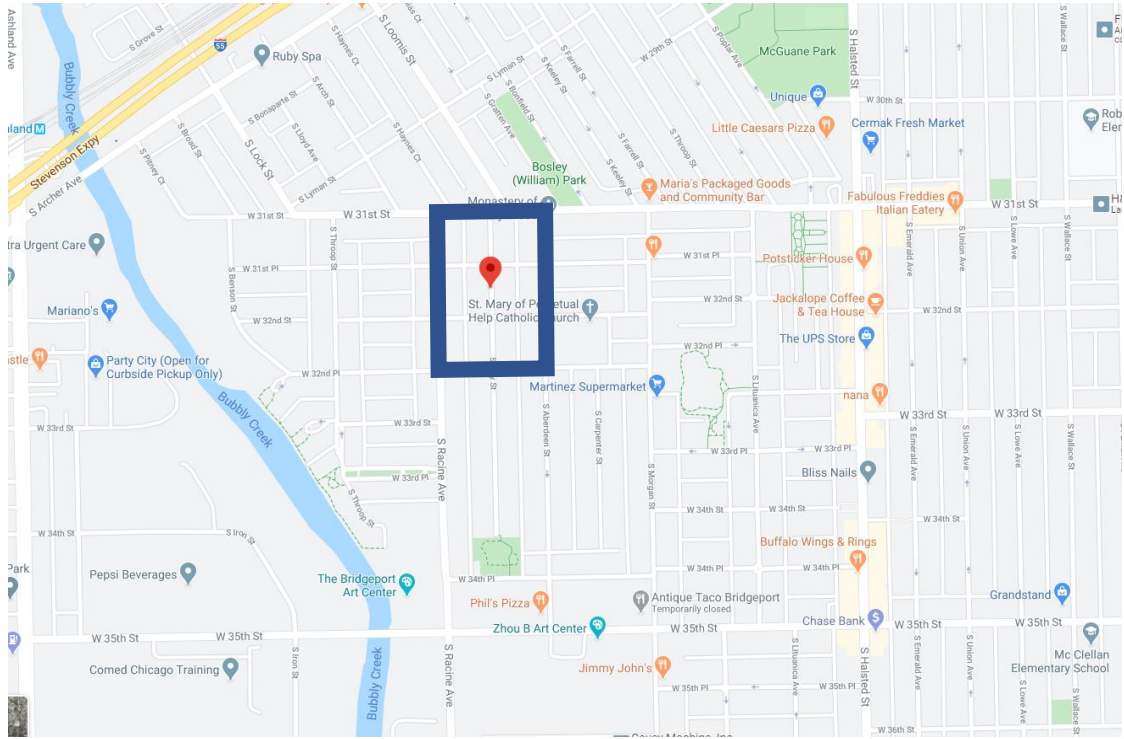

**Figure 8 The map of the Chicago case**

**Table 5. The entropy of CSE and GW modified index for the Chicago case**

|              | Retail | Office | Industry | Service | Entertainment | Residential | Entropy |
|--------------|--------|--------|----------|---------|---------------|-------------|---------|
| Original CSE | 0.0%   | 0.0%   | 2.2%     | 0.0%    | 4.3%          | 93.5%       | **0.16** |
| +GW          | 1.3%   | 4.5%   | 5.5%     | 8.1%    | 4.3%          | 76.2%       | **0.53** |
| +GW+RSW      | 5.8%   | 16.3%  | 10.3%    | 9.8%    | 20.4%         | 37.4%       | **0.90** |

These two cases show that for a single measurement index, the new measures based on GW and RSW significant improve the validity of the entropy measure compared to the original CSE, as they more accurately reflect true land use mix.

## 4.2 Cluster Analysis

For the cluster analysis, as shown in the table 6, each of the 10 groups identified has unique features based on land use proportion. The characteristics of land use, entropy, and dependent variables are summarized in Table 5, showing the unique attributes for each group, sorted by descending entropy index order with GW and RSW. For comparison, the statistics of the national average are also listed at the end of Table 5.

Table 6. Summary statistics of each land use composition group

| Gp | Land use | | | | | | Entropy | | |
|---|---|---|---|---|---|---|---|---|---|
| | Ret | Off | Ind | Svc | Ent | Res | CSE | +GW | +GW +RSW |
| 1 | 1% | 1% | 2% | 3% | 1% | 91% | 0.18 | 0.21 | 0.50 |
| 2 | 4% | 4% | 52% | 12% | 3% | 25% | 0.65 | 0.65 | 0.63 |
| 3 | 3% | 5% | 5% | 62% | 4% | 20% | 0.55 | 0.56 | 0.64 |
| 4 | 4% | 39% | 8% | 18% | 6% | 25% | 0.73 | 0.74 | 0.65 |
| 5 | 22% | 6% | 7% | 16% | 20% | 29% | 0.81 | 0.82 | 0.72 |
| 6 | 3% | 3% | 28% | 8% | 3% | 55% | 0.60 | 0.61 | 0.74 |
| 7 | 3% | 2% | 6% | 8% | 3% | 77% | 0.38 | 0.44 | 0.75 |
| 8 | 12% | 6% | 5% | 11% | 11% | 56% | 0.66 | 0.70 | 0.78 |
| 9 | 3% | 3% | 4% | 22% | 3% | 65% | 0.50 | 0.56 | 0.80 |
| 10 | 5% | 5% | 7% | 35% | 5% | 43% | 0.67 | 0.70 | 0.82 |
| Average* | 6% | 7% | 12% | 20% | 5% | 50% | 0.44 | 0.48 | 0.68 |

*Average summary of the mean value of each land use ratio and entropy measure

From the results, each of the groups has its own unique features in land use proportion. Some of the groups are mainly composed of single land use compared with the national average. For instance, Group 1 is mainly residential, Group 2 has significantly higher industry use, and Group 3 is dominated by service land use. Some of the compositions are more complex. More than one land use is substantially higher/lower than the national average. For instance, Groups 5 and 8 have higher retail and entertainment, though Group 5 has much lower residential use.

The changes to the entropy index between the original and new indexes are inconsistent between each group, which is mainly the result of RSW. As shown in the table 6, the national average land use composition is 6%, 7%, 12%, 20%, 5%, and 50% for retail, office, industry, service, entertainment and residential land use, respectively, which is far from an equal proportion of 16.67% for every category. Therefore, the groups with a land use composition more similar to an equal ratio than the national average get strongly punished by the new RSW method, as seen in Groups 2, 4, and 5, where the index value becomes lower than CSE. Moreover, GW generally increases the entropy values among all the land use groups, which is also expected since the consideration of surrounding areas will more likely introduce more diversified land use into the equation.

Therefore, the cluster analysis result in table 6 shows that the new measures better reflect the sites' actual land use mix by revealing more detail about the proportion and combination of the mix instead of a single dimension index based on the binary spectrum.

## 4.3 Performance on the Regression

Table 7 The performance of entropies in regression results

| Housing price | | | |
|---|---:|---:|---:|
| **CSE** | 10933.0 | 4.6 | 0.000 |
| **+GW** | 92952.0 | 37.5 | 0.007 |
| **+GW & RSW** | 145390.0 | 0.5 | 0.000 |
| **GW & Cluster*** | 186048 ~ 298122 | 12.6 ~ 29.20 | 0.018 |
| **Drive alone** | | | |
| **CSE** | 0.01 | 3.306 | 0.000 |
| **+GW** | -0.14 | 90.35 | 0.037 |
| **+GW & RSW** | 10.0 | 17.195 | 0.001 |
| **GW & Cluster*** | 0.70 ~ 0.80 | 12.6 ~ 29.20 | 0.036 |
| **Tweet density (log)** | | | |
| **CSE** | 2.89805 | 112.2 | 0.056 |
| **+GW** | 4.81974 | 185.8 | 0.141 |
| **+GW & RSW** | 4.6921 | 151.5 | 0.098 |
| **GW & Cluster*** | -3.64 ~ 0.07 | 70.1 ~ 107.5 | 0.156 |

*As the cluster method is a multi-measure path, its coefficients need to consider the intercept coefficient, which indicates the base group. Therefore, the coefficients are not directly comparable with the other single measure methods.

Unlike the case study results, further quantitative regression analysis reveals that the improvements from both GW and RSW are not very consistent across different research subjects. All the results that strongly contradict with the previous studies are highlighted in the table 7.

For housing price, better land use mix is supposed to increase the available facilities and public services around the site, which should be reflected in higher housing prices based what homebuyers' willingness like to pay (Song & Knaap, 2004). All three new indexes and CSE show a positive coefficient in the results, which means housing price is expected to increase with the increase of land use mix. In comparison, the multi-measure clustering methods show more variety in that different cluster types have different levels of influence on housing price. At the same time, the interpretation of multi-measure indexes coefficients do not indicate a direct change, but rather the expected difference compared with the base group, which makes the interpretation of the results much less intuitive than the single measure entropies.

For t-value, the GW & RSW combined entropy index is not significantly associated with housing price, though all the other methods show significant t-values, especially GW. R2 also varies quite a bit between each method; both CSE and the GW & RSW method have extremely small R2 values, meaning the relationship is very weak. This renders the new measures weak when predicting housing price. The clustering method has the largest R2 at 0.018, which is more than twice that of the second-highest entropy, GW. Therefore, in general, the clustering method has the most reliable performance given all its results have the largest R2 value, and all the clusters show significant relationships with housing price.

For the percentage of people that drive alone to commute, the new methods' performances are similarly complicated as for the housing price. In theory, a higher land use mix should increase the available job locations around the site within walking distance, thus decreasing the expected value of the drive alone ratio. However, both CSE and the GW & RSW combined measure yield significant positive coefficients, indicating the drive alone ratio is expected to rise with increased land use mix. Still, the GW & RSW result especially does not make sense, as only a 0.1 increase in land use entropy would expect to increase the drive alone ratio by 100%. The R2 results further prove the invalidity of both CSE and the GW & RSW method, which have extremely small R2 values that indicate trivial explanatory power in the model. In general, both GW and the multi-measure clustering method have the strongest performances that show valid coefficient, t-value, and R2 results.

For tweet density, land use mix should provide more locations around the site for residents to conduct social and recreational activities, therefore increasing the likelihood of sending geo-tagged tweets to share their lives. The results of all the measures demonstrate the expected positive coefficient direction with very significant associations. However, the R2 results do differ, with the multi-measure clustering method having the highest value at 0.156, meaning nearly 16% of tweet density variance can be explained by this measure. The second highest R2 value is the GW-only entropy measure at 0.141, and CSE follows at 0.056. In general, both the clustering and GW measures perform very well for all indicators.

## 5. Discussion and Conclusion

### 5.1 A Very Lucky Coincidence

To sum, GW was demonstrated to improve the entropy measure from both a quantitative and qualitative perspective. But, unexpectedly, we found the entropy measure with RSW did not have the expected excellent performance in the regression, as it did in the case study comparison with the GW-only and multi-measure clustering methods. To better explain this, in table 8 we summarized the average value of housing price, percentage of drive-alone commuting, and tweet density in each cluster group.

Table 8. The land use pattern groups and their performance of dependent variables

|  | Land use | | | | | | Entropy | | | Dp variables | | |
|---|---|---|---|---|---|---|---|---|---|---|---|---|
| Gp | Ret | Off | Ind | Svc | Ent | Res | CSE | +GW | +GW +RSW | Pr | Dri | Tw |
| 1 | 1% | 1% | 2% | 3% | 1% | 91% | 0.18 | 0.21 | 0.5 | 237k | 0.8 | 0.5 |
| 2 | 4% | 4% | 52% | 12% | 3% | 25% | 0.65 | 0.65 | 0.63 | 189k | 0.76 | 1.5 |
| 3 | 3% | 5% | 5% | 62% | 4% | 20% | 0.55 | 0.56 | 0.64 | 277k | 0.7 | 11 |
| 4 | 4% | 39% | 8% | 18% | 6% | 25% | 0.73 | 0.74 | 0.65 | 288k | 0.69 | 27 |
| 5 | 22% | 6% | 7% | 16% | 20% | 29% | 0.81 | 0.82 | 0.72 | 298k | 0.74 | 14.2 |
| 6 | 3% | 3% | 28% | 8% | 3% | 55% | 0.6 | 0.61 | 0.74 | 186k | 0.79 | 0.7 |
| 7 | 3% | 2% | 6% | 8% | 3% | 77% | 0.38 | 0.44 | 0.75 | 271k | 0.75 | 1.8 |
| 8 | 12% | 6% | 5% | 11% | 11% | 56% | 0.66 | 0.7 | 0.78 | 293k | 0.73 | 6.5 |
| 9 | 3% | 3% | 4% | 22% | 3% | 65% | 0.5 | 0.56 | 0.8 | 294k | 0.72 | 2.9 |
| 10 | 5% | 5% | 7% | 35% | 5% | 43% | 0.67 | 0.7 | 0.82 | 278k | 0.73 | 5.5 |
| Average* | 6% | 7% | 12% | 20% | 5% | 50% | 0.44 | 0.48 | 0.68 | 259k | 0.76 | 3.7 |

*Average summary of the mean value of each land use ratio and entropy measures

The table 7 is sorted by GW & RSW entropy values. It shows that the entropy values of Groups 3, 4, and 5 stood out as significant outliers in the association with the dependent variables. These groups' GW & RSW entropy values were significantly punished by RSW compared to CSE and the GW-only measure due to their specialized land use pattern, especially low residential use feature. As for the dependent variables, these three groups had the highest average tweet density and lowest drive-alone commute ratio. The housing price level was also not low for them. But why did Groups 3, 4, and 5, with their fairly clustered land use patterns, have such good performance, even better than the highly mixed land use pattern groups? The reasons behind this performance are that Group 3's high service land use usually represented major hospitals or schools; Group 4's high office land use typically represented the downtown area; and Group 5's high retail and entertainment land use generally indicated tourist destinations or urban entertainment centers. Therefore, it is understandable that these groups performed well with these dependent variables.

On the other hand, Group 6 was also an outlier, where even its GW & RSW entropy was higher than Groups 3, 4, and 5. However, its housing price and tweet density were much lower than these three groups, and its drive-alone commuting ratio was higher. This association conflicts should also be cause by the Group 6's land use pattern composition. Even though Group 6 has balanced residential use, its highly clustered industry land use indicated it is possibly in an industrial zone, which makes its weak performance understandable.

Therefore, it seems that there is a very lucky coincidence regarding the influence of land use mix patterns. Even the equal proportion of land use assumption in CSE, which does not have any theoretical foundation compared to RSW, which is based on complete community theory, coincidently performs better in certain predictive aspects, such as housing price, commuting mode, and tweet density, as tested in this paper. This is because, for many places such as downtown areas, major service facilities, or entertainment centers that we usually illustrate as having good urban life performance with high land use mix, land use for office, service, entertainment, and retail are expected to cluster compared to regional land use structures. Even residential land use also exists in these areas, its proportion is highly likely to be lower than the regional average. Therefore, thanks to this lucky coincidence, the equal, CSE-based proportion has been considered as somewhat valid (Ewing & Cervero, 2010) in many extant empirical studies, especially in the transportation field,. This is also why we found the new index measures performed weaker after we fixed its MAUP and context issues.

The more robust performance does not mean we agree that the arbitrary equal proportion assumption used in the original CSE is right. The nature of the city decides that not all land use will be equal in general, so an equal proportion of land use would mean some land use type is over-clustered in a given area. Over-clustered land use in one region would inevitably deprive the possibility of other regions having a similar level of land use. Therefore, this equal proportion assumption is, in essence, unfeasible to implement across the city and, more importantly, goes against the original spirit of the land use mix concept (Jacobs, 1961), which encourages instead of removes the possibility of creating more areas with various land use in close proximity to one another. We believe that land use mix based on complete community theory to consider regional context is the most feasible way to implement this concept.

Even the new index showed that the associations between land use mix and its impacts are not as strong as we expected, We still believe it does not conflict with early quantitative works that propose and promote land use mix. Most of the developments that Jacobs (1961) opposed using the land use mix concept are mainly single-use, high-density residential projects, such as Pruitt-Igoe in New York, which were related to the urban renewal movement and modernist architecture paradigm at the time. In this analysis, such developments could belong to the Group 1 type of land use pattern. The result also shows that Group 1 had a much lower housing price, a higher ratio of driving alone to commute, and lower tweet density compared with the other grups. This fits with what early studies (e.g., Alexander, 1977; Jacobs, 1961) pointed out. When looking at the results from Group 1 versus the rest of the groups, the land use mix theories still hold. Therefore, it is important to note that instead of rejecting the role of land use mix, the findings in this paper deepen the understanding of it by exploring more detailed land use combination patterns.

Based on this discussion, we believe that the GW & RSW measure's weak performance, as well as the clustering multi-measure method's consistent, good performance, have the same conclusion: with the increasing availability of data, a single measure can no longer sufficiently represent complex land use patterns in space to test its influence. A single measure can only be built based on a single definition of a "good" mix, with a linear spectrum to describe the level of "goodness." However, as the cluster analysis in this study showed, there are multiple different land use combination patterns that can achieve "good" economic, transportation, or social influence. Even RSW is based on complete community theory, though it remains unable to explain every advantage of all land use patterns, such as downtown areas' performances. Therefore, the multi-measure clustering method is much better at addressing land use pattern complexity when compared with single measure methods.

## 5.2 Limitations and Future Direction

Given all the measures' algorithms, the quality of the final product is dependent on the quality of its data source. In this paper, the land use mix measure was compiled from employment data from the LEHD dataset instead of direct land use parcel data. Therefore, there could be a potential difference when this method is applied to land use parcel data, which might influence the results. Additionally, as shown in the LEHD document, the data is subject to non-sampling errors, including misreported data, late reporters whose records are missing or imputed, and geographic/industry edits and imputations. We do argue that employment data has been applied in many studies before to measure land use, proving to be valid and reliable as proxy data. The LEHD dataset is also the only (current) dataset available to us to represent national land use at the census block group level. On the other hand, the quantitative studies cited in the paper only used basic linear regression to test the performance in the association, which limits this study's further implications for causation inferences.

Moreover, even though the multi-measure method had the most robust performance of the new measures, it also has its limitations. The interpretation of clustering groups is much less intuitive than the single 0-1 entropy index measure because the regression result coefficient has to consider the base group and intercept to fully express effect size in each group.

For future research, we believe that there are two new directions the new measures proposed in this study can extend to. First, the LEHD dataset we used to measure land use is actually a longitudinal dataset that offers data from 2002 to 2017, which leaves us great potential to analyze the change dynamics of land use mix in the U.S. over a 10-year range to see if recent U.S. urban development policies are making changes to the land use mix situation. The other direction is applying the new algorithms in other study fields that concern the mix of a different group of components, such as measuring the combination of different income-level residents or the mix of different racial groups in a city to examine the change dynamic of social equity in space.

## 5.3 Conclusion

To sum, from the analysis, we found that the GW, RSW, and clustering multi-measure methods are all able to fix the validity issues of the CSE measure for land use mix. For the single index methods, GW proved to substantially improve both the validity and predictive power of the land use mix measure compared to CSE by addressing the MAUP and considering not only the given site, but also its surrounding areas. Related, the results of the GW & RSW index were more complicated, as discussed above. The case study indicated RSW does help better reflect the ground truth of land use mix on-site, as it considers regional land use structures instead of using equal portions as the best-case scenario. But, in the quantitative analysis, RSW showed lesser predictive power regarding land use mix by showing irregular coefficients or extremely small R2 values. Detailed cluster analysis revealed that the conflict is caused by "a very luck coincidence." Even the equal proportion assumption has no theoretical grounds compared with RSW, as CSE's equal proportion scenario fits better in the analysis, especially when explaining places like downtown, schools, or entertainment centers, where the proportion of each land use is more equalized instead of similar to the regional land use structure. Due to this complexity of land use composition patterns and their associations with various subjects, the multi-measure method showed the most consistently robust performance compared with the GW or GW & RSW single entropy measures.

In short, we conclude that the clustering multi-measure method is the first recommended choice for researchers because it allows the model to consider land use pattern complexity. If the study requires more simplicity or intuitive interpretations of its results, the GW-based entropy measure is the ideal choice to reflect land use mixture. If a researcher has a clear definition of land use mix based on complete community theory or a similar paradigm, we recommend the GW & RSW entropy measure to most accurately measure the concept they define.

The reason behind this "coincidence" raises another important question that has not been clearly answered for years regarding the land use mix concept: is there a single, universal best composition for land use mix across the world? The urban study field has long discovered that compared with having only a single residential use area, proximity to other types of land use is necessary. But what is the best land use pattern, and what is the best ratio for each land use type? Prior studies have not given enough attention to unveiling this mystery fully. From the results of this paper, we believe that using one single-dimension definition to define *mixed* or *not mixed* is very difficult, especially when studying the empirical relationship between land use mix and its influence. For different influence subjects, the "best" land use composition varies, and there could also be multiple "best" land use compositions existing at the same time, such as with our Groups 3, 4, and 5 all having good performances in commuting methods and tweet density. Therefore, we suggest moving the topic of measuring land use mix to the next level by measuring land use pattern. Land use pattern is a multi-dimensional concept that allows richer and more colorful variation in land use combinations instead of one universal *mix* definition.